\documentclass[11pt]{article}
\pdfoutput=1
\usepackage{graphicx}
\usepackage{epsfig}
\usepackage{amsfonts}
\usepackage{amscd}
\usepackage{latexsym}
\usepackage{amsmath,amssymb}
\usepackage{verbatim}
\usepackage{setspace}
\usepackage{color}
\usepackage{fancyhdr}
\usepackage{cite}
\usepackage{hyperref}
\usepackage{tensor}
\usepackage{xfrac}

\usepackage[textheight=9in, textwidth=6.5in, letterpaper]{geometry}

\numberwithin{equation}{section}

\def\ve{{\varepsilon}}

 \def\p{\partial}
 \def\bz{{\bar z}}
 
\def\0{{(0)}}
\def\1{{(1)}}
\def\2{{(2)}}

\def\co{{\cal O}}

\def\ci{{\mathcal I}}

\def\<{\langle }
\def\>{\rangle }

\def \eps {\epsilon}
\def\bz{{\bar z}} \def\zb{{\bar z}}

\def\p{\partial}

\def\ci {{\cal I }}

\def\CW{{\cal W}}

\def\zb {{\bar{z}}}

\newcommand{\bea}{\begin{eqnarray}}
\newcommand{\eea}{\end{eqnarray}}
\newcommand{\be}{\begin{equation}}
\newcommand{\ee}{\end{equation}}
\newcommand{\ba}{\begin{aligned}}
\newcommand{\ea}{\end{aligned}}

\def\be{\begin{equation}}
\def\ee{\end{equation}}
\def\beq{\be\begin{array}{c}}
\def\eeq{\end{array}\ee}

   \makeatletter
  \let\over=\@@over \let\overwithdelims=\@@overwithdelims
  \let\atop=\@@atop \let\atopwithdelims=\@@atopwithdelims
  \let\above=\@@above \let\abovewithdelims=\@@abovewithdelims
\renewcommand\section{\@startsection {section}{1}{\z@}%
                                   {-3.5ex \@plus -1ex \@minus -.2ex}
                                   {2.3ex \@plus.2ex}%
                                   {\normalfont\large\bfseries}}

\renewcommand\subsection{\@startsection{subsection}{2}{\z@}%
                                     {-3.25ex\@plus -1ex \@minus -.2ex}%
                                     {1.5ex \@plus .2ex}%
                                     {\normalfont\bfseries}}

\newcommand{\SNcheck}[1]{{\color{purple}\checkmark}}
\newcommand{\MPcheck}[1]{{\color{red}\checkmark}}
\newcommand{\NPcheck}[1]{{\color{magenta}\checkmark}}

\usepackage{tikz}
\usepackage{slashed}
\usetikzlibrary{calc}   
\usetikzlibrary{topaths}
\usetikzlibrary{decorations}
\usetikzlibrary{decorations.pathmorphing}

{\cfoot{\thepage}}
\begin{document}
\begin{titlepage}
\unitlength = 1mm~\\
\vskip 3cm
\begin{center}

{\LARGE{\textsc{The Soft $\mathcal{S}$-Matrix in Gravity}}}

\vspace{0.8cm}
Elizabeth Himwich{$^{*}$}, Sruthi A. Narayanan{$^{*}$}, Monica Pate{$^{*\dagger}$}, \\
Nisarga Paul{$^{*}$}, and Andrew Strominger{$^{*}$}\\
\vspace{1cm}

{$^*$\it  Center for the Fundamental Laws of Nature, Harvard University,\\
Cambridge, MA 02138, USA} \\
{$^\dagger$\it Society of Fellows, Harvard University, Cambridge, MA 02138, USA}

\vspace{0.8cm}

\begin{abstract}

\end{abstract}

\end{center}
The gravitational $\mathcal{S}$-matrix  defined with an infrared (IR) cutoff factorizes  into hard and soft factors. 
The soft factor is universal and contains all the IR and collinear divergences. Here we show, in a momentum space basis, that the
intricate expression for the soft factor is fully reproduced by two boundary currents, which live on the celestial sphere. The first  of these is the supertranslation current, which generates spacetime supertranslations. The second is its symplectic partner, the Goldstone current for spontaneously broken supertranslations. The current algebra has an off-diagonal level structure involving the gravitational cusp anomalous dimension and the logarithm of the IR  cutoff. It is further shown that the gravitational memory effect is contained as an IR safe observable within the soft ${\mathcal S}$-matrix. 

\end{titlepage}

\tableofcontents

\section{Introduction}

The scattering amplitudes of four-dimensional asymptotically flat quantum gravity transform covariantly under asymptotic symmetries, which include the infinite-dimensional symmetry group of a two-dimensional conformal field theory \cite{Cachazo:2014fwa,Kapec:2014opa}.  In particular, $SL(2,\mathbb{C})$ Lorentz transformations act as the global conformal symmetry on the celestial sphere $\mathcal{CS}$ at null infinity.  This observation lends scattering amplitudes a natural reinterpretation as correlation functions of a holographically dual ``celestial conformal field theory" (CCFT) on the celestial sphere
\cite{He:2015zea, Strominger:2017zoo}:
\begin{equation} \label{eq:map}
\langle \text{out} | \mathcal{S} | \text{in} \rangle \rightarrow \langle \mathcal{O}_1 \cdots \mathcal{O}_n  \rangle_{\mathcal{CS}}.
\end{equation}
The asymptotic symmetries  additionally include supertranslations, whose associated conservation
laws are  equivalent to Weinberg's soft graviton theorem \cite{Strominger:2013jfa,He:2014laa}. In $U(1)$ gauge theory, the soft photon theorem is the conservation laws following from the asymptotic large gauge symmetries \cite{He:2014cra,Campiglia:2015qka,Kapec:2015ena}. 
 
These symmetries are realized  as  current algebras on the celestial sphere. This structure both governs the form of and explains the origin of infrared (IR) divergences: the latter  are needed in order to set to zero all conservation-law-violating exclusive amplitudes \cite{Kapec:2017tkm, Choi:2017bna,Choi:2017ylo}.

Soft factorization is a closely related  universal IR feature of scattering amplitudes.  It posits that scattering amplitudes (with an IR cutoff) can be factored into hard and soft pieces, where the soft piece is given by a universal formula that contains all IR divergences of the full amplitude. This factorization was shown in \cite{Nande:2017dba} for QED to be celestially realized as the familiar factorization of 2D CFT  correlations functions into a universal ``conformally soft" current algebra factor and a theory-dependent ``conformally hard" factor. The full soft $\mathcal{S}$-matrix was reproduced from a 2D current algebra on the celestial sphere. 

The purpose of this paper is to extend the  analysis of \cite{Nande:2017dba} to  gravitational theories. We determine, in a momentum space basis,  the current algebra on the celestial sphere that governs the leading IR behavior of gravitational scattering amplitudes.  In addition to the previously-studied
 supertranslation current \cite{Strominger:2013jfa,He:2014laa}, we find that the celestial algebra includes a second ``Goldstone current" that is the gradient of the Goldstone mode  associated to  spontaneously broken supertranslation symmetry. It has a  non-vanishing level proportional to the gravitational cusp anomalous dimension and the logarithm of the IR cutoff. The soft part of gravitational scattering amplitudes is shown to be entirely reproduced by current algebra correlation functions.   In direct analogy with the gauge theory analysis, the soft piece is a correlation function of Wilson line-inspired operators constructed from the Goldstone boson and supertranslation currents. Gravitational Wilson line operators and their relation to gravitational soft dressings were discussed in \cite{Choi:2017ylo,Choi:2019fuq,White:2011}.
 The gravitational cusp anomalous dimension, which appears in the level of the Goldstone current, was related to   soft factorization  in~\cite{Miller:2012an}.

 Generic scattering processes induce transitions between inequivalent vacua, which can be measured through the gravitational memory effect\cite{Strominger:2014pwa}. This effect is encoded in the soft ${\mathcal S}$-matrix. 
 Inequivalent vacua are related by supertranslations and the memory effect measures the transformation of the Goldstone boson under the symmetry. Memory  is an IR safe observable and we show it is reproduced in the current algebra by the IR finite operator product expansion (OPE) between the supertranslation current and Goldstone boson.

The paper is organized as follows.  Section \ref{sec:pre} presents notation and conventions. In Section \ref{sec:weinberg}, we review the relation  between supertranslation symmetry and IR behavior of gravitational scattering matrix elements.  In Section \ref{sec:currentall}, we recast the statements from Section \ref{sec:weinberg} as  current algebra identities on the celestial sphere. In Subsection \ref{sec:Wilson} we introduce, for hard massless external particles, Wilson line-like operators   constructed from the Goldstone mode. In Subsection \ref{sec:current2}, we review the soft graviton mode, which realizes the supertranslation  current on the celestial sphere, and derive the OPEs of the associated algebra.  In Section \ref{sec:softfact}, we turn to IR divergences in scattering amplitudes arising from internal soft graviton exchanges.  We  review the  Feynman diagrammatic formula for the soft $\mathcal{S}$-matrix and demonstrate that it is reproduced by correlation functions of the Goldstone mode.  In Subsection \ref{sec:gravity}, we show that the soft $\mathcal{S}$-matrix for an arbitrary number of soft gravitons and hard particles is reproduced by correlators involving both  supertranslation currents and Goldstone bosons. In Subsection \ref{sec:memory}, we relate the OPEs determined previously to the gravitational memory effect. In Section \ref{sec:appendix}, by modifying the Wilson line operators,
we generalize the analysis in Section \ref{sec:softfact} to include external massive particles. An extension of this analysis to a conformal primary basis will be included in \cite{Arkani-Hamed:2020gyp}.

\section{Preliminaries}

\label{sec:pre}

We consider four-dimensional perturbative gravity coupled to massless matter on a Minkowski background. We parametrize null momenta as
\begin{equation} \label{eq:momentum}
    p^\mu_k = \eta_k \omega_k \hat q^\mu (z_k, \bz_k) 
\end{equation}
where $\omega_k\geq 0$, $\eta_k = \pm 1$ for outgoing and incoming particles respectively, and 
\begin{equation}
    \hat{q}^\mu(z,\bar{z}) = (1+z\bar{z}, z+\bar{z}, -i(z-\bar{z}),1-z\bar{z}),\quad \quad \quad  \hat{q}^\mu(z_i,\bar{z}_i)\hat{q}_\mu(z_j , \bz_j) = -2|z_i-z_j|^2 \equiv -2 |z_{ij}|^2.
\end{equation}
Polarization tensors for spinning massless particles are constructed from the polarization vectors
\begin{equation}
     \ve_{k+}^\mu = \frac{1}{\sqrt{2}} \p_{z_k} \hat q^\mu (z_k, \bz_k), \quad \quad \quad 
     \ve_{k-}^\mu = \frac{1}{\sqrt{2}} \p_{\bz_k} \hat q^\mu (z_k, \bz_k).
\end{equation}
The polarization tensor for a positive helicity outgoing graviton is $\ve_{k+}^{\mu\nu} = \ve_{k+}^\mu \ve_{k+}^\nu$.

We employ flat Bondi  coordinates $(u, r, z, \bz)$, in which the standard Cartesian coordinates are
\begin{equation}
    x^\mu = \frac{1}{2}\left( u n^\mu+   r \hat q^\mu (z, \bz)\right), 
\end{equation} 
where 
\begin{equation} \label{defn}
n^\mu = (1,0,0,-1), \ \ \  n^\mu\hat{q}_\mu(z,\bar{z}) = -2.
\end{equation}
The Minkowski line element becomes 
\begin{equation}
    ds^2 = \eta_{\mu\nu}dx^{\mu}dx^{\nu} = -   dudr +    r^2 dz d\bz.
\end{equation}
In these coordinates, the celestial sphere $\mathcal{CS}$ is conformally mapped to a plane on which $(z_k, \bz_k)$ labels the point where a massless particle with momentum $p_k$ crosses null infinity $\mathcal{I}$.

\section{Supertranslations and Weinberg's soft graviton theorem}
\label{sec:weinberg}

In this section, we review the equivalence between Weinberg's soft graviton theorem and supertranslation symmetry of the $\mathcal{S}$-matrix.

Introducing the soft graviton mode operators $P_z^\pm$~\cite{He:2014laa,Strominger:2013jfa}, defined as
\begin{equation} \label{eq:softgrav1}
    \begin{split}
     \kappa   P_z^+  & = -\lim_{\omega \to 0} 
      \p_\bz   \left (\omega a^{\rm out}_+( \omega, z, \bz) +\omega  a^{\dagger \rm out}_-( \omega, z, \bz) \right), \\
     \kappa P_z^-  & = -\lim_{\omega \to 0}  
        \p_ \bz \left (\omega a^{\rm in}_+( \omega, z, \bz) +\omega  a^{\dagger\rm in}_-  ( \omega, z, \bz) \right), 
    \end{split}
\end{equation}
 where $\kappa = \sqrt{32\pi G}$, Weinberg's soft graviton theorem for a scattering process involving $n$ hard massless particles and one soft graviton of momentum $q^\mu = \omega \hat q^\mu (z, \bz)$ and polarization $\ve^{\mu\nu}$ becomes 
\begin{equation}
\label{eq:PSoft1}
    \begin{split}
        \langle \mbox{out}|P_z^+\mathcal{S} - \mathcal{S}P_z^-|\mbox{in}\rangle
            &= -\lim_{\omega\rightarrow 0}\partial_{\bar{z}}\left[\omega\sum_{k=1}^n\frac{ \ve^+_{\mu\nu}p_{k}^\mu p_{k}^\nu}{p_{k}\cdot q}\right]\langle \mbox{out}|\mathcal{S}|\mbox{in}\rangle
            =   \sum_{k=1}^n\frac{\eta_k \omega_k}{z-z_k} \langle\mbox{out}|\mathcal{S}|\mbox{in}\rangle.
    \end{split} 
\end{equation} 
A Fourier mode decomposition of the graviton field in turn relates the soft graviton mode $P_z^+$ on ${\mathcal I}^+$ and boundary components of the asymptotic metric 
\begin{equation} \label{eq:softgrav}
    \begin{split}
     4 G   P_z^+  & = \int_{-\infty}^\infty du ~\p_\bz  N_{zz} = \p_\bz C_{zz}|_{\ci^+_+} - \p_\bz C_{zz}|_{\ci^+_-},
    \end{split}
\end{equation}
where $N_{zz} = \p_u C_{zz}$ is the radiative component of the gravitational field (see \cite{Strominger:2017zoo} and references therein for definitions).  Similar expressions hold for $P_z^-$. 

The statement of supertranslation invariance of the gravitational $\mathcal{S}$-matrix is
\begin{equation}
    Q^+ \mathcal{S} - \mathcal{S} Q^- = 0, 
\end{equation} 
which follows from taking the derivative of \eqref{eq:PSoft1} with respect to $\bar{z}$ and then integrating the result against an arbitrary function $f(z,\bar{z})$: 
\be
    \begin{split}
\label{eq:MassivePWard33}
\int \frac{d^2z}{2\pi} f(z,\bar{z})\partial_{\bar{z}}\langle \mbox{out}|P_z^+\mathcal{S} - \mathcal{S}P_z^-|\mbox{in}\rangle & = \int d^2z f(z,\bar{z})\sum_{k=1}^n \eta_k \omega_k\delta^{(2)}(z_k-z)\langle \mbox{out}|\mathcal{S}|\mbox{in}\rangle\\
 &\equiv\langle \mbox{out}|Q^+_H \mathcal{S} - \mathcal{S} Q^-_H|\mbox{in}\rangle.
 \end{split}
\ee
The hard charges $Q_H^\pm$ implement the action of the supertranslation symmetry on matter.   A single hard massless asymptotic state transforms as\footnote{The operator equation for a classical symmetry $\phi \rightarrow \phi + \epsilon \delta \phi$ with charge $Q$ is  $\delta \mathcal{O} = i \left[Q, \mathcal{O}\right]$.}
 \begin{equation}\label{eq:Infinitesimal35}
    \delta_f | p_k \rangle =i Q_H^\pm | p_k \rangle= i \eta_k\omega_kf(z_k,\bar{z}_k)| p_k \rangle
 \end{equation}
where $Q^\pm_H$ acts on states with $\eta_k = \pm1$, respectively. For the full charge $Q^\pm = Q^\pm_H + Q^\pm_S$, the left-hand side of \eqref{eq:MassivePWard33} is identified as $-\langle \mbox{out}|Q^+_S \mathcal{S} - \mathcal{S} Q^-_S|\mbox{in}\rangle$. The soft charges $Q^\pm_S$ act by adding soft gravitons, performing a non-trivial vacuum transformation.  

Supertranslation symmetry is spontaneously broken in the standard Minkowski vacuum, giving rise to a Goldstone boson,  denoted $C$. $C$ is canonically paired with the soft graviton $P_z^+$  and  related to a boundary component of the asymptotic metric \cite{He:2014laa}:
\be
    C_{zz}|_{\mathcal{I}^+_-}=-\partial_z^2 C.
\ee
The Goldstone boson transforms under an infinitesimal supertranslation by an inhomogeneous shift:
\be \label{Ctrans}
    \delta_f C =f. 
\ee

\section{Currents on the celestial sphere}
\label{sec:currentall}
In this section, we recast the supertranslation symmetry of $\mathcal{S}$-matrix elements presented in Section \ref{sec:weinberg} as the action of a current algebra on the celestial sphere.

\subsection{Gravitational Wilson lines}
\label{sec:Wilson}

Weinberg's soft graviton theorem, as reviewed in the previous section, implies that asymptotic states of hard massless particles transform non-trivially under supertranslations. Here, we recast the action of supertranslations on celestial correlators as the transformation of Wilson line-inspired operators constructed from gravitational boundary modes, which we therefore also include in the supertranslation current algebra. 

    Massless particles in momentum eigenstates are associated to unique points on $\mathcal{CS}$ and are represented by local operators $\co_k (p_k)$, which transform under supertranslations according to \eqref{eq:Infinitesimal35}: 
\be
\label{eq:InfinitesimalOp}
\delta_f\mathcal{O}_k(p_k) = i\eta_k\omega_kf(z_k,\bar{z}_k)\mathcal{O}_k(p_k).
\ee   
 To isolate the supertranslation transformation properties of $\co_k$, we introduce operators of the form\footnote{Note that the argument of the exponent is dimensionless because $\left[C\right] = \left[\frac{1}{r} g_{zz}\right] = \left[\text{mass}\right] ^{-1}$.}
\be
\label{eq:WilsonLine1}  \mathcal{W}_k(p_k) =  e^{i\eta_k \omega_k C(z_k,\bar{z}_k)},
\ee   
and decompose
\be \label{eq:decomp}
    \mathcal{O}_k = \mathcal{W}_k\tilde{\mathcal{O}}_k.
\ee
The transformation of $\mathcal{W}_k$ follows from \eqref{Ctrans},
\be
\delta_f\mathcal{W}_k(z_k,\bar{z}_k) = i \eta_k \omega_k f(z_k,\bar{z}_k)\mathcal{W}_k(z_k,\bar{z}_k),
\ee
and accounts for the full transformation of $\mathcal{O}_k(p_k)$ in \eqref{eq:InfinitesimalOp}, implying  that $\tilde{\mathcal{O}}_k$ is invariant under supertranslations. Note that neither $\CW_k$ nor $\tilde \co_k$ alone create physical scattering states. Physical states are dual to the composite operator $\co_k$.

\subsection{Summary of current algebra}
\label{sec:current2}

On the celestial sphere, Weinberg's soft graviton theorem can be recast as the insertion of a current 
\begin{equation}
     P_z \equiv P_z^+- P_z^-.
\end{equation}
Using \eqref{eq:map}, the soft theorem \eqref{eq:PSoft1} becomes
\begin{equation}
    \langle P_z  \mathcal{O}_1 \cdots \mathcal{O}_n\rangle
         =  \sum_{k=1}^n\frac{\eta_k \omega_k}{z-z_k}  \langle \mathcal{O}_1 \cdots \mathcal{O}_n\rangle, 
\end{equation}
which is immediately recognized as the Ward identity of a Kac-Moody symmetry generated by $P_z$.

The decomposition \eqref{eq:decomp} implies that $P_z$ and $\mathcal{W}_k$ have the OPE
\begin{equation} \label{eq:PW}
    P_z\mathcal{W}_k\sim \frac{\eta_k \omega_k}{z-z_k}\mathcal{W}_k.
\end{equation}
Using the expression \eqref{eq:WilsonLine1} for $\mathcal{W}_k$ in terms of the Goldstone operator $C$ then gives 
\begin{equation} \label{eq:PzC}
    P_z C(w,\bar{w}) \sim \frac{- i}{z-w}.
\end{equation}
This OPE is of the general form allowed by $SL(2,\mathbb{C})$ conformal symmetry, under which $C$ has conformal weight  $(-\frac{1}{2},- \frac{1}{2})$.  The weight of $C$ follows from that fact that the argument of the exponential in \eqref{eq:WilsonLine1} must have net zero conformal weight and that energies $\omega_k$ transform with conformal weight $(\frac{1}{2}, \frac{1}{2})$.  As shown in \cite{Donnay:2018neh}, $P_z$ is a spin-one operator of conformal dimension $\left(\frac{3}{2}, \frac{1}{2}\right)$, and its action raises the conformal dimension of operators $\mathcal{O}_k$ by  $\left(\frac{1}{2},\frac{1}{2}\right)$.

To deduce the OPE of two supertranslation currents, note that soft gravitons have vanishing energy and do not couple at leading order in a low-energy expansion \cite{Weinberg:1965nx}, implying that  
\begin{equation}\label{eq:PP}
P_zP_w \sim 0.
\end{equation}
In analogy with the supertranslation current, we define a Goldstone current
\begin{equation} \label{eq:deptildeP}
\tilde{P}_z = i\partial_z C, 
\end{equation}
and from \eqref{eq:PzC}, we immediately find the OPE
\begin{equation}
P_z\tilde{P}_w \sim \frac{1}{(z-w)^2}.
\end{equation} 
The dimension of $C$ implies that $\tilde{P}_z$ has dimension $\left(\frac{1}{2},-\frac{1}{2}\right)$, so we expect an OPE of the form
\begin{equation} \label{eq:tildeP2}
\tilde{P}_z \tilde{P}_w \sim k\frac{(\bar{z}-\bar{w})}{(z-w)},
\end{equation}
where $k$ is a constant that we will determine in the following section (see equation \eqref{eq:kvalue} therein).

\section{Infrared divergences from virtual gravitons}
    \label{sec:softfact}

In addition to singularities arising from the emission and absorption of soft gravitons, which are captured by the soft graviton theorem, scattering amplitudes in four-dimensional theories of gravity contain IR divergences arising from virtual soft gravitons exchanged between external legs.    As Weinberg first explained \cite{Weinberg:1965nx}, virtual graviton exchange contributes a universal soft factor to the  $\mathcal{S}$-matrix. This section reproduces the virtual soft exchange factor in the $\mathcal{CS}$ current algebra.  
    
We begin by recalling the general formula for IR divergences derived in e.g. \cite{Naculich:2011ry}.  The $\mathcal{S}$-matrix for scattering $n$ hard massless particles factorizes as
    \begin{equation} \label{eq:sf1}
        \langle {\rm out} | \mathcal{S} | {\rm in } \rangle =
            \exp\left[\frac{1}{\epsilon}\frac{G}{2\pi}\sum_{i,j=1}^n p_i \cdot p_j \ln\left( \frac{2p_i \cdot p_j}{\mu^2}\right)\right]
             \widehat {\langle {\rm out} | \mathcal{S}  | {\rm in } \rangle} 
    \end{equation}
in $d = 4+2\epsilon$ dimensional regularization\footnote{When the IR regulator is a cutoff $\lambda_{IR}$, the exponent becomes proportional to $\log(\lambda_{IR})$ rather than $\eps^{-1}$.} with $\mu$ an arbitrary mass scale and $ \widehat {\langle {\rm out} | \mathcal{S}  | {\rm in } \rangle}$ IR finite. 
    
    Working with the momentum parametrization \eqref{eq:momentum}, we find
    \begin{equation}
        \begin{split}
            \sum_{i,j=1}^n p_i \cdot p_j \ln\left( \frac{2p_i \cdot p_j}{\mu^2}\right)
               & = \sum_{i,j=1}^n p_i \cdot p_j \ln\left( \frac{- 4\eta_i \eta_j \omega_i \omega_j |z_{ij}|^2}{\mu^2}\right)  = \sum_{i\neq j}^n p_i \cdot p_j  \ln |z_{ij}|^2 ,
        \end{split}
    \end{equation}
    where the last equality follows from expanding the logarithm into a sum of terms, using $p_i^2=0$, and invoking total momentum conservation to obtain identities such as
    \begin{equation}
        \sum_{i,j=1}^n p_i \cdot p_j  \ln (\eta_j \omega_j)
             =  \sum_{j=1}^n \ln (\eta_j \omega_j) p_j^\mu  \left (\sum_{i = 1}^n p_{i\mu}\right) = 0.
    \end{equation} 
    With these simplifications, \eqref{eq:sf1} becomes
     \begin{equation} \label{eq:sf2}
        \langle {\rm out} | \mathcal{S} | {\rm in } \rangle =
            \exp\left[-\frac{1}{\epsilon}\frac{G}{\pi}\sum_{i\neq j}^n  \eta_i \eta_j \omega_i \omega_j|z_{ij}|^2 \ln |z_{ij}|^2\right]
             \widehat {\langle {\rm out} | \mathcal{S}  | {\rm in } \rangle}.
    \end{equation}
    
    To recast this result on the $\mathcal{CS}$, note that the decomposition \eqref{eq:decomp} of operators according to their supertranslation symmetry transformation properties implies factorization of correlation functions:
    \begin{equation}\label{eq:corrfactor}
        \langle\mathcal{O}_1\cdots \mathcal{O}_n\rangle = \langle \mathcal{W}_1 \cdots \mathcal{W}_n\rangle\langle\tilde{\mathcal{O}}_1 \cdots \tilde{\mathcal{O}}_n\rangle.
    \end{equation}
    The $\mathcal{W}_k$ and $\tilde {\mathcal{O}}_k$ correlators  are the  soft (IR divergent) and hard (IR finite) factors, respectively.  
    
    It follows that the general form of the IR divergent piece in \eqref{eq:sf2}, involving a sum over pairs $i,j$, can be reproduced by a correlation function of exponential operators $\mathcal{W}_k$ for which the two-point function is the only non-vanishing connected $n$-point correlation function of the exponentiated field $C$.  Thus expressing the correlation function of $\mathcal{W}_k$ operators in terms of correlation functions of $C$, we find
    \begin{equation} \label{eq:Wsimp}
        \langle \mathcal{W}_1 \cdots \mathcal{W}_n\rangle = 
            \exp \left [- \frac{1}{2}\sum_{i \neq  j}^n \eta_i \eta_j \omega_i \omega_j \langle C(z_i, \bz_i)C(z_j, \bz_j) \rangle \right].
    \end{equation}
    Comparing with \eqref{eq:sf2}, we immediately deduce
   \begin{equation} \label{eq:C2point}
        \langle C(z_i,\bar{z}_i)C(z_j,\bar{z}_j)\rangle = \frac{1}{\eps} \frac{2G}{\pi                       }|z_{ij}|^2\ln |z_{ij}|^2, 
    \end{equation}
    from which, using the definition \eqref{eq:deptildeP} of $\tilde{P}_z$, we fix the level of the Goldstone current algebra $k$ in \eqref{eq:tildeP2} to be
    \begin{equation} \label{eq:kvalue}
        k = \frac{1}{\eps} \frac{2G}{\pi}.
    \end{equation}
    While the level of the Goldstone current is IR divergent, we will show in the next section that it does not affect IR safe observables such as gravitational memory.

\section{Current algebra correlators and the soft $\mathcal{S}$-matrix}

Having fixed the leading singularities in the OPEs of the  soft photon currents $P_z$ and Goldstone operators $C$, we now show how these results can be used to calculate the soft part of the $\mathcal{S}$-matrix in gravity, including the emission of soft gravitons and the gravitational memory effect. 

\subsection{Gravity} \label{sec:gravity}

A scattering process involving $m$ soft gravitons and $n$ hard massless particles
is represented by a celestial correlation function, which factors according to the decomposition \eqref{eq:decomp}:
\begin{equation} 
    \langle  P_{z_1} \cdots P_{z_m} \mathcal{O}_1 \cdots \mathcal{O}_n \rangle  
        = \langle  P_{z_1} \cdots P_{z_m} \mathcal{W}_1 \cdots \mathcal{W}_n \rangle  
            \langle   \tilde{\mathcal{O}}_1 \cdots \tilde{\mathcal{O}}_n \rangle .
\end{equation}
Now, we will use the current algebra specified in Sections 
\ref{sec:currentall} and \ref{sec:softfact} to compute the soft part of the $\mathcal{S}$-matrix given by the correlation function of soft photon currents $P_z$ and operators $\mathcal{W}_k$.

If we assume that $P_z$ is holomorphic away from other operator insertions,
subleading terms in its OPE with other operators cannot contribute to correlation functions. 
Then, the leading singularities given by \eqref{eq:PW}
and \eqref{eq:PP} imply a further factorization: 
\begin{equation}\label{eq:int1}
  \langle P_{z_1} \cdots P_{z_m} \mathcal{W}_1 \cdots \mathcal{W}_n \rangle = \left[\prod_{i=1}^{m} \sum_{j=1}^n \frac{\eta_j \omega_j}{z_i - z_j} \right] \langle  \mathcal{W}_1 \cdots \mathcal{W}_n \rangle, 
\end{equation}
which is simply the manifestation of Weinberg's soft graviton theorem on the celestial sphere.

Finally, combining \eqref{eq:Wsimp} and \eqref{eq:C2point}, we find 
\begin{equation}
  \langle P_{z_1} \cdots P_{z_m} \mathcal{W}_1 \cdots \mathcal{W}_n \rangle = \left[\prod_{i=1}^{m}\sum_{j=1}^n \frac{\eta_j \omega_j}{z_i - z_j} \right] 
    \exp \left [- \frac{1}{\eps} \frac{G}{\pi}\sum_{k \neq  \ell}^n \eta_k \eta_\ell \omega_k \omega_\ell  
    |z_{k \ell}|^2\ln |z_{k \ell}|^2 \right].
\end{equation}

\subsection{Gravitational memory}
\label{sec:memory}

In this subsection, we recompute the gravitational memory effect from the current algebra. The key point is that gravitational memory involves $P_z\tilde P_w$, not $\tilde P_z \tilde P_w$, and also only involves ratios of amplitudes. It is therefore an IR safe observable.

We begin by focusing on \eqref{eq:int1} for a single insertion of $P_z$ and noting that it can be written as 
\begin{equation}
    \langle P_{z}   \mathcal{W}_1 \cdots \mathcal{W}_n \rangle
    =  \left [  \sum_{j=1}^n  i \eta_j \omega_j \langle P_z C(z_j, \bz_j)\rangle \right]  \langle  \mathcal{W}_1 \cdots \mathcal{W}_n \rangle.
\end{equation}
Rearranging, we find 
\begin{equation} \label{eq:ratio1}
    \sum_{j=1}^n  i \eta_j \omega_j \langle P_z C(z_j, \bz_j)\rangle 
         = \frac{\langle P_{z}   \mathcal{W}_1 \cdots \mathcal{W}_n \rangle}{ \langle  \mathcal{W}_1 \cdots \mathcal{W}_n \rangle} = \frac{\langle P_{z}   \mathcal{O}_1 \cdots \mathcal{O}_n \rangle}{ \langle  \mathcal{O}_1 \cdots \mathcal{O}_n \rangle},
\end{equation}
where we use the factorization property of correlation functions \eqref{eq:corrfactor} to obtain the last equality.

We next use crossing symmetry \cite{Strominger:2013jfa}
\begin{equation}
    \langle {\rm out }  |
    P_z^+ \mathcal{S}|{\rm in}\rangle 
    = -   \langle {\rm out }  | \mathcal{S}
    P_z^-|{\rm in}\rangle 
\end{equation}
to  relate this ratio of correlation functions to a scattering amplitude with a single insertion of $P_z^+$: 
\begin{equation}
    \frac{1}{2}\frac{\langle P_{z}   \mathcal{O}_1 \cdots \mathcal{O}_n \rangle}{ \langle  \mathcal{O}_1 \cdots \mathcal{O}_n \rangle}
     =\frac{ \langle {\rm out }  |
    P_z^+ \mathcal{S}|{\rm in}\rangle }{ \langle {\rm out }  |
    \mathcal{S}|{\rm in}\rangle }.
\end{equation} 
In turn, the right-hand side can be interpreted as the expectation value of the change in the asymptotic metric $\Delta h_{\mu\nu}$ induced by the scattering of $n$ hard particles \cite{Strominger:2014pwa}, where $h_{\mu\nu}$ is the perturbation of the  metric  about Minkowski space
\begin{equation}
    g_{\mu\nu} = \eta_{\mu\nu}+ \kappa h_{\mu\nu}. 
\end{equation}   
Altogether, we find 
\begin{equation}
   \sqrt{\frac{2\pi}{G}} \p_\bz \left ( \lim_{r \to \infty }r \ve_+^{\mu\nu} \Delta h_{\mu\nu}(r,z,\zb ) \right) 
         =  \frac{\langle P_{z}   \mathcal{O}_1 \cdots \mathcal{O}_n \rangle}{ \langle  \mathcal{O}_1 \cdots \mathcal{O}_n \rangle}.
\end{equation}
Using \eqref{eq:ratio1} and 
\begin{equation}
     i \eta_j \omega_j \langle P_z C(z_j, \bz_j)\rangle 
         = \frac{\eta_j \omega_j }{z-z_j} =  -\partial_{\bar{z}}\left[ \frac{ \ve^+_{\mu\nu}p_{j}^\mu p_{j}^\nu}{p_{j}\cdot \hat q (z, \bz)}\right], 
\end{equation}
we obtain the Braginsky-Thorne formula \cite{BraginskyThorne} for gravitational memory due to the scattering of massive bodies: 
\begin{equation}
       \lim_{r \to \infty }r \ve_+^{\mu\nu} \Delta h_{\mu\nu}(r,z,\zb)  
          = -\sqrt{\frac{G}{2\pi}} \sum_{j = 1}^n\frac{ \ve^+_{\mu\nu}p_{j}^\mu p_{j}^\nu}{p_{j}\cdot \hat q(z,\zb) }.
\end{equation}
The gravitational memory formula is thus determined by the $P_z C$ OPE. Note that unlike $\langle CC\rangle$, $\langle P_z C\rangle$ is IR finite and directly related to an IR safe observable: \eqref{eq:ratio1} involves a \emph{ratio} of scattering amplitudes that precisely cancels the IR divergences due to virtual gravitons.

\section{Soft $\mathcal{S}$-matrix for massive particles}\label{sec:appendix}

In this section, we show that the Goldstone two-point function derived above from the soft $\mathcal{S}$-matrix for massless external particles correctly reproduces the soft $\mathcal{S}$-matrix for massive external particles. 

Massive particles in momentum eigenstates are not canonically associated to points on the celestial sphere, but instead to points on a resolution of timelike infinity by a three-dimensional hyperboloid. We thus use coordinates that give a hyperbolic slicing of 
Minkowski space:
\begin{equation}
    x^\mu = \frac{\tau}{2\rho}(n^\mu+\rho^2\hat{q}^\mu(z,\bar{z})), 
\end{equation}
in which the line element becomes
\begin{equation}
    ds^2 = -d\tau^2+\tau^2\left(\frac{d\rho^2}{\rho^2}+\rho^2 dz d\bar{z}\right)
    .
\end{equation} 
Surfaces of constant $\tau$ are constant curvature hyperbolic slices and their $\rho=\infty$ boundary is the celestial sphere, labelled by points $(z,\bar{z})$. Massive momenta are naturally parametrized by points $(\rho_k, z_k, \bz_k)$ on the hyperboloid, defined by the mass-shell condition $p_k^2 =-m_k^2$: 
\begin{equation} \label{massivepar}
p_k^\mu = \frac{\eta_km_k}{2\rho_k}(1+\rho_k^2(1+z_k\bar{z}_k), \rho_k^2(z_k+\bar{z}_k), -i\rho_k^2(z_k-\bar{z}_k), -1+\rho_k^2(1-z_k\bar{z}_k)) \equiv \eta_km_k \hat p_k^\mu. 
\end{equation}
Massive particles with  momenta $p_k$  approach the points $(\rho_k, z_k, \bz_k)$ on the hyperboloid at timelike infinity ($\tau \to \pm \infty$). We will also make use of the following inner products:
\begin{equation}
    \begin{split}
        \hat p_k \cdot \hat q(z, \bz) & = - \left (\frac{1}{\rho_k}   + \rho_k |z- z_k|^2 \right), \\ \quad \quad \quad 
        \hat p_k \cdot \hat p_\ell& = - \frac{1}{2} \left ( \frac{\rho_k}{\rho_\ell} + \frac{\rho_\ell}{\rho_k} + \rho_k \rho_\ell |z_{k\ell}|^2  \right)
        \equiv - \cosh \gamma_{k \ell},
    \end{split} 
\end{equation}
where $\gamma_{k \ell}$ is implicitly defined to be real for any pair (outgoing or incoming) of on-shell momenta.

To generalize the Wilson line operators to massive particles, we begin by rewriting \eqref{eq:WilsonLine1} as
\begin{equation} \label{genWilson1}
    \mathcal{W}_k (p_k) =\exp \left[-i \int \frac{d^2 z}{2  \pi } C(z, \bz)~ \p_{\bar{z}}^2 \left(\frac{\ve^+_{\mu\nu} p_k^\mu p_k^\nu} {p_k\cdot \hat q(z, \bz) }\right)\right].
\end{equation} 
Using the parametrization \eqref{massivepar} of massive momenta, 
 \eqref{genWilson1} becomes
\be
\mathcal{W}_k(p_k) = \exp\left[\frac{i\eta_km_k}{2}\int d^2z\mathcal{G}^{(3)}(\hat{p}_k;z,\bar{z})C(z,\bar{z})\right] 
\ee
where the Green's functions $\mathcal{G}^{(n)}$ are given generally by \cite{Campiglia:2015lxa} 
\be
\mathcal{G}^{(n)}(\hat{p}_k;z,\bar{z}) = \frac{(n-1)}{2\pi}\frac{\rho_k^n}{(1+\rho_k^2|z-z_k|^2)^n}. 
\ee

The $n$-point correlation function of such operators takes the form 
\begin{equation} \label{eq:w2point}
\langle\mathcal{W}_1\cdots\mathcal{W}_n\rangle = \exp\left[-\frac{1}{8}\sum_{i,j=1}^n\eta_i\eta_jm_im_j\int d^2w d^2z\mathcal{G}^{(3)}( \hat p_i;w,\bar{w})\mathcal{G}^{(3)}(\hat p_j;z,\bar{z})\langle C(w,\bar{w})C(z,\bar{z})\rangle\right],
\end{equation} 
where here the sum includes $i=j$ terms, which arise from  Wick contractions within a single $\mathcal{W}_k$.\footnote{Note that in the massless analysis, we do not encounter such terms because the two-point function of $C$ vanishes as the points are taken to be coincident. This reflects the well-known fact that collinear divergences cancel in gravity \cite{Weinberg:1965nx}.} 
Explicitly evaluating this expression using the two-point function \eqref{eq:C2point} determined previously, we find 

 \begin{equation} \label{finalexp1} \begin{split}
\langle \mathcal{W}_1\cdots\mathcal{W}_n\rangle =  \exp&\left[-\frac{1}{\epsilon}\frac{G}{4\pi}\sum_{i,j = 1}^n   \eta_i\eta_jm_im_j  \right.
\\& \quad \quad  \times  \left. \left(\gamma_{ij}\left[\frac{1}{\sinh^2\gamma_{ij}}+ {4\cosh^2\gamma_{ij}}  \right]^{\frac{1}{2}}-2\ln(\rho_i\rho_j)\cosh\gamma_{ij}+\frac{\rho_i}{\rho_j}+\frac{\rho_j}{\rho_i}\right)\right].
\end{split}
\end{equation}  Invoking total momentum conservation simplifies the result, as in Section 5. Using, in particular, 
\begin{equation}
    \begin{split}
    \sum_{i,j = 1}^n\eta_i\eta_jm_im_j
        \ln(\rho_i\rho_j)\cosh\gamma_{ij}
             = -2  \sum_{i,j = 1}^n  \ln(\rho_i) p_i \cdot p_j
                = -2  \sum_{i = 1}^n  \ln(\rho_i) p_i^\mu  \sum_{j=1}^n  p_{j\mu}&=0,\\
     \sum_{i,j = 1}^n\eta_i\eta_jm_im_j \left (\frac{\rho_i}{\rho_j}+\frac{\rho_j}{\rho_i}\right) 
     =2\sum_{i=1}^n \frac{\eta_i m_i}{\rho_i}
     \sum_{j = 1}^n \eta_j m_j \rho_j 
      =-2\sum_{i=1}^n \frac{\eta_i m_i}{\rho_i}
      n^\mu  \sum_{j = 1}^n p_{j \mu} & = 0,
      \end{split}
\end{equation}
where $n$ is the constant null vector defined in \eqref{defn},  \eqref{finalexp1} simplifies to 
\begin{equation} \label{eq:massiveW}
\langle\mathcal{W}_1\cdots\mathcal{W}_n\rangle= \exp\left[-\frac{1}{\epsilon}\frac{G}{4\pi} \sum_{i, j=1}^n\eta_i\eta_jm_im_j\gamma_{ij}\left(\frac{1}{\sinh^2\gamma_{ij}}+ {4\cosh^2\gamma_{ij}}  \right)^{\frac{1}{2}}\right].
\end{equation}
This correlation function reproduces the soft factor of the $\mathcal{S}$-matrix, shown in~\cite{Weinberg:1965nx,Weinberg:1995mt} to be 
\be\label{fieldtheory}
\langle {\rm out} | \mathcal{S} | {\rm in } \rangle =  \exp\left[-\frac{1}{\epsilon}\frac{G}{4\pi}\sum_{i, j=1}^n\eta_i\eta_j m_im_j(\gamma_{ij}-i\pi\delta_{\eta_i\eta_j,1})\left(\frac{1}{\sinh^2\gamma_{ij}}+4\cosh^2\gamma_{ij}\right)^{\frac{1}{2}}\right] \widehat {\langle {\rm out} | \mathcal{S}  | {\rm in } \rangle}, 
\ee 
provided that when $\eta_i \eta_j= 1$, we analytically continue $\gamma_{ij} \rightarrow \gamma_{ij}-i\pi$ in \eqref{eq:massiveW}.  

\section*{Acknowledgements}

We are grateful to Nima Arkani-Hamed, Aditya Parikh, and Ana Raclariu for useful discussions.  This work was supported by DOE grant DE-SC/0007870 and by the Gordon and Betty Moore Foundation and John Templeton Foundation grants via the Black Hole Initiative.
N.P. was supported by the Purcell fellowship.
M.P. acknowledges
the support of a Junior Fellowship at the Harvard Society of Fellows.

\bibliography{softsmatrix}
\bibliographystyle{utphys}
\end{document}